\documentclass[twocolumn,showpacs,preprintnumbers,amsmath,amssymb]{revtex4}

\usepackage{graphicx}
\usepackage{dcolumn}
\usepackage{bm}

\newcommand{\be}{\begin{equation}}
\newcommand{\ee}{\end{equation}}
\newcommand{\bea}{\begin{eqnarray}}
\newcommand{\eea}{\end{eqnarray}}
\newcommand{\ba}{\begin{eqnarray}}
\newcommand{\ea}{\end{eqnarray}}

\newcommand{\beq}{\begin{equation}}
\newcommand{\eeq}{\end{equation}}
\newcommand{\beqa}{\begin{eqnarray}}
\newcommand{\eeqa}{\end{eqnarray}}
\newcommand{\beqar}{\begin{eqnarray*}}
\newcommand{\eeqar}{\end{eqnarray*}}

\newcommand{\reef}[1]{(\ref{#1})}

\newcommand{\ie}{{\it i.e.,}\ }

\newcommand{\labell}[1]{\label{#1}} 
\newcommand{\etal}{{\it et al.}}

\newcommand{\X}{\mathcal{X}}
\newcommand{\Z}{\mathcal{Z}}
\newcommand{\fin}{f_\infty}
\newcommand{\lp}{\ell_{\mt P}}

\def\t6 {T_\mt{D6}}

\newcommand{\mt}[1]{\textrm{\tiny #1}}
\newcommand{\ads}{a_d^*}
\newcommand{\pe}{\rho_\mt{E}}
\newcommand{\hC}{{\widehat C}}

\begin{document}

\preprint{arXiv:1006.1263 [hep-th]}

\title{Seeing a c-theorem
with holography}

\author{Robert C. Myers and Aninda Sinha}
 \affiliation{Perimeter Institute for Theoretical Physics, Waterloo,
Ontario N2J 2Y5, Canada }


\begin{abstract}
There is no known model in holography exhibiting a $c$-theorem where
the central charges of the dual CFT are distinct. We examine a
holographic model of RG flows in a framework where the bulk gravity
theory contains higher curvature terms. The latter allows us to
distinguish the flow of the central charges $a$ and $c$ in the dual
field theories in four dimensions. One finds that the flow of $a$ is
naturally monotonic but that of $c$ is not. Extending the analysis of
holographic RG flows to higher dimensions, we are led to formulate a
novel c-theorem in arbitrary dimensions for a universal coefficient
appearing in the entanglement entropy of the fixed point CFT's.

\end{abstract}
\maketitle

\noindent {\bf Introduction:} Zamolodchikov's c-theorem \cite{zam} is a
remarkable result for quantum field theories in $d=2$. A direct outcome
of the c-theorem is that in any renormalization group (RG) flow
connecting two fixed points,
\be (c)_{\rm UV}\ge (c)_{\rm IR}\,. \labell{beta} \ee
That is, the central charge of conformal field theory (CFT) describing
the ultraviolet fixed point is larger than (or equal to) that at the
infrared fixed point. The proof relies only on the euclidian group of
symmetries, the existence of a conserved stress-energy tensor and
unitarity in the field theory.

There have been various suggestions on how such a result might extend
to quantum field theories in higher $d$. Cardy \cite{cardy} conjectured
a monotonic flow for the coefficient of the A-type trace anomaly. For
even $d$, the trace anomaly for a CFT is in a curved background is
given by \cite{traca}
\be \langle\,T^a{}_a\,\rangle = \sum B_i\,I_i -2\,(-)^{d/2}A\, E_d
\labell{trace} \ee
where $E_d$ is the Euler density in $d$ dimensions and $I_i$ are the
independent Weyl invariants of weight $-d$ \cite{convention}. Cardy's
proposal is then that $A$ satisfies a relation like eq.~\reef{beta}
along RG flows between two fixed points in any even $d$.
%
%
Of course, this coincides with Zamolodchikov's result in $d=2$ where
$A= c/12$.

Efforts made towards proving Cardy's conjectured c-theorem have focused
$d=4$ in which case eq.~\reef{trace} contains two terms, \ie there is a
single Weyl invariant $I_1=C_{abcd}C^{abcd}$. The common nomenclature
denotes the two central charges as $c=16\pi^2B_1$ and $a= A$.
%
%
Numerous nontrivial examples have been found supporting Cardy's
conjecture, including perturbative fixed points \cite{jack} and
supersymmetric gauge theories \cite{anselmi}. The latter investigations
also demonstrated that the $d=4$ central charge $c$
will not generically satisfy eq.~\reef{beta}. Further, as we review
below, support for a c-theorem in higher dimensions was found
\cite{gubser} with the AdS/CFT correspondence \cite{revue}.

However, a general proof is lacking and, in fact, a counter-example to
Cardy's c-theorem in $d=4$ was proposed in \cite{yuji}. If this
counter-example survives further scrutiny, there are two obvious
possibilities: A `c-theorem' exists in higher $d$ but the quantity
satisfying eq.~\reef{beta} is not the central charge $A$.
Alternatively, the central charge $A$ satisfies eq.~\reef{beta} under
RG flows in higher $d$ but only when some more stringent requirements
are imposed than in $d=2$. In the latter case, the challenge would be
to identify the precise conditions for which the analog of
eq.~\reef{beta} is satisfied in higher (even) $d$.

In the following, we find evidence for the second alternative within
the framework of the AdS/CFT correspondence. We examine the c-theorem
using a holographic model with a higher curvature gravity theory
\cite{old1,old2} which allows one to distinguish the central charges
$a$ and $c$ in the dual CFT. Hence we are able to discriminate between
the behaviour of these two central charges in RG flows and we find that
only $a$ has a natural monotonic flow. We are also able to extend our
analysis to holographic CFT's in arbitrary higher $d$. Our results
there suggest the following general conjecture:
 \vskip 3pt
\noindent
\it Placing a $d$-dimensional CFT on $S^{d-1}\times R$ and
    calculating the entanglement entropy of the ground state
    between two halves of the sphere, one finds a universal
    contribution: $S_{univ}\propto a^*_d$ (as detailed in eq.~\reef{unis}).
    Further in RG flows between fixed points, $(a^*_d)_{\rm
    UV}\ge(a^*_d)_{\rm IR}$.
 \vskip 3pt
\noindent \rm This conjecture then gives us a framework in which to
consider the c-theorem for $d$ even or odd. As described below, this
conjecture actually coincides to Cardy's proposal for even $d$. We now
discuss the holographic origin of this conjecture.

\noindent {\bf Holographic c-theorem:} The AdS/CFT correspondence has
emerged as a powerful tool to study the behaviour of strongly coupled
CFT's in diverse dimensions \cite{revue}. Within this framework,
\cite{gubser} considered the c-theorem where one begins with
($d$+1)-dimensional Einstein gravity coupled to various matter fields:
\be I=\frac{1}{2\lp^{d-1}}\int d^{d+1} x \, \sqrt{-g} \left(R+
{\mathcal L}_{\rm matter}\right) \labell{action} \ee
The matter theory is assumed to have various stationary points where
${\mathcal L}_{\rm matter}={d(d-1)\alpha_i^2/L^2}$ with some canonical
scale $L$. The fixed points are distinguished by different values of
$\alpha$ as indicated by the subscript and at these points, the gravity
vacuum is simply AdS$_{d+1}$ with the curvature scale given by $\tilde
L^2=L^2/\alpha_i^2$. RG flows between critical points can be described
with a metric of the form
\be ds^2=e^{2 A(r)}\left( -dt^2+ d\vec{x}_{d-1}^2 \right)+dr^2\,.
\labell{metric} \ee
This metric becomes that for AdS$_{d+1}$ with $A(r)=r/\tilde L$ at the
stationary points. Now define \cite{gubser}:
\be a(r)\equiv\frac{\pi^{d/2}}{\Gamma\left(d/2\right)\left(\lp
A'(r)\right)^{d-1}}\,, \labell{def0} \ee
where `prime' denotes a derivative with respect to $r$. Then for
general solutions of the form \reef{metric}, one finds
\bea a'(r)&=&-\frac{(d-1)\pi^{d/2}}{\Gamma\left(d/2\right)\lp^{d-1}
A'(r)^d} A''(r)
\labell{magic0}\\
&=& -\frac{\pi^{d/2}}{\Gamma\left(d/2\right)\lp^{d-1}
A'(r)^d}\left(T^t{}_t-T^r{}_r\right) \ge0\,.\nonumber
 \eea
Above in the second equality, the Einstein equations are used to
eliminate $A''(r)$ in favour of components of the stress tensor. The
final inequality assumes that the matter fields obey the null energy
condition \cite{HE}. Given this monotonic evolution of $a(r)$ with $r$
and the standard connection between $r$ and energy scale in the CFT,
$a(r)$ always decreases in flowing from the UV to the IR.

To make better contact with the dual CFT, it is simplest to focus the
discussion on $d=4$ at this point. Then with the holographic trace
anomaly \cite{sken} for the AdS$_5$ stationary points, one finds
\be a(r)\big|_{AdS} =\pi^2\,{\tilde L^3}/{\lp^3}=a\,. \labell{acharge}
\ee
That is, the value of the flow function \reef{def0} matches precisely
that of the central charge $a$ in the dual CFT at each of the fixed
points. Hence with the assumption of the null energy condition, the
holographic CFT's dual to the gravity theory \reef{action} satisfy
Cardy's c-theorem. Of course, one must add that the two central charges
are precisely equal \cite{sken}, \ie $a=c$, for the class of $d=4$
CFT's dual to Einstein gravity. Hence these holographic models do not
distinguish between the flow of $a$ and $c$.

It has long been known that to construct a holographic model where
$a\ne c$, the gravity action must include higher curvature interactions
\cite{highc}. In part, this motivated the construction of
quasi-topological gravity \cite{old1}
 \be
I=\frac {1}{2\lp^3}\int d^5x \sqrt{-g}
\left[\frac{12\alpha^2}{L^2}+R+\frac{\lambda}2 L^2 \X_4+\frac{7\mu}4
L^4 \Z_5\right]
 \labell{qtaction}
 \ee
where $\X_4$ is the `Gauss-Bonnet' interaction, a combination of
curvature-squared terms which corresponds to the Euler density in four
dimensions, and $\Z_5$ is a particular interaction involving
curvature-cubed terms developed in \cite{old1,chile}. This holographic
model allows one to explore the full three-parameter space of
coefficients controlling the two- and three-point functions of the
stress tensor in a general four-dimensional CFT \cite{osborn}. However,
one must keep in mind that this action \reef{qtaction} was not derived
from string theory. Rather it was constructed to facilitate the
exploration of a broader class of holographic CFT's while maintaining
control within the gravity calculations. Further, the gravitational
couplings in eq.~\reef{qtaction} can be constrained by various
consistency requirements of the dual CFT \cite{old2}. A generalization
of the model \reef{qtaction} to $d=2$ CFT's and c-theorems was
considered in \cite{sinha}.

In comparison to the action presented in \cite{old1}, we have replaced
the cosmological constant term by $12\alpha^2/L^2$ with the study of RG
flows in mind. The idea is that as in eq.~\reef{action} the gravity
theory is coupled to a standard matter action with various stationary
points which yield different values for the parameter $\alpha^2$. The
curvature scale of the AdS$_5$ vacua are related to the scale $L$
appearing in eq.~\reef{qtaction} by $\tilde L^2=L^2/\fin$ where
\be \alpha^2-\fin+\lambda\,\fin^2+\mu\,\fin^2=0\,. \labell{cubic} \ee
Using standard techniques \cite{sken}, one determines the central
charges of the CFT's dual to these AdS$_5$ vacua \cite{old2}:
\bea a&=& \pi^2\tilde L^3/\lp^3\, \left(1-6\lambda f_\infty +9\mu
f_\infty^2 \right)\,,\labell{aa}\\
c&=& \pi^2\tilde L^3/\lp^3\, \left(1-2\lambda f_\infty -3\mu f_\infty^2
\right)\,.\labell{cc} \eea

Considering the metric ansatz \reef{metric}, it is straightforward to
construct two flow functions:
\bea a(r)&\equiv&{\pi^2\over \lp^{3} A'(r)^{3}}\, \left(1-{6}\lambda
L^2A'(r)^2+ {9}\mu L^4A'(r)^4 \right)\ \labell{afun}\\
c(r)&\equiv&{\pi^2\over \lp^{3} A'(r)^{3}}\, \left(1-{2}\lambda
L^2A'(r)^2- {3}\mu L^4A'(r)^4 \right)\ \ \labell{cfun} \eea
These are chosen as the simplest extensions of eq.~\reef{def0} with
$d=4$ which yield the two central charges at the fixed points, \ie
$a(r)|_{AdS}=a$ and $c(r)|_{AdS}=c$. Now for a general RG flow
solution, one finds
\be a'(r)=-3\frac{A''(r)}{A'(r)}\, c(r)= -\frac{\pi^2}{\lp^3
A'(r)^4}\left(T^t{}_t-T^r{}_r\right)\ge0\,, \labell{magic1} \ee
where as before we are using the gravitational equations of motion and
assuming that the matter sector obeys the null energy condition. Hence
with the latter assumption, $a(r)$ evolves monotonically along the
holographic RG flows and we can conclude that the central charge
\reef{aa} satisfies  the analog of eq.~\reef{beta}. One can also
consider the behaviour of $c(r)$ along RG flows but there is no clear
way to establish that $c'(r)$ has a definite sign. Hence this
holographic model provides another broad class of four-dimensional
theories which support Cardy's proposal that the central $a$ (rather
than any other central charge) evolves monotonically along RG flows.

Given this result and those above for Einstein gravity, it is
straightforward to extend our holographic analysis of quasi-topological
gravity to an arbitrary spacetime dimension. Beginning with the
equations of motion which are proportional to $T^t{}_t-T^r{}_r$, one
can engineer the following flow function \cite{avatar}
\bea a_d(r)&\equiv&{\pi^{d/2}\over\Gamma\left(d/2\right)
\left(\lp A'(r)\right)^{d-1}}\, \labell{adfun}\\
&&\times \
 \left(1-{2(d-1)\over
d-3}\lambda L^2A'(r)^2
- {3(d-1)\over d-5}\mu L^4A'(r)^4 \right)\,. \nonumber \eea
By construction, $a_d(r)$ satisfies the following:
\be a'_d(r)=-\frac{\pi^{d/2}}{\Gamma\left(d/2\right)\lp^{d-1}
A'(r)^{d}}\left(T^t{}_t-T^r{}_r\right)\ge0 \labell{magic2} \ee
where we again assume the null energy condition. Note that here we must
also ensure that $A'(r)>0$ for odd $d$ -- the details of this proof
will be given in \cite{future}. If we define the fixed point value as
$\ads\equiv a_d(r)|_{AdS}$ then
\be \ads={\pi^{d/2}\tilde L^{d-1}\over\Gamma\left(d/2\right)\lp^{d-1}}
\left(1-{2(d-1)\over d-3}\lambda \fin - {3(d-1)\over d-5}\mu \fin^2
\right)\,, \labell{astar}\ee
and the result in eq.~\reef{magic2} guarantees that
\be \left(\ads\right)_{UV}\ge \left(\ads\right)_{IR}\,. \labell{beta3}
\ee

Having found that $\ads$ satisfies a c-theorem, one is left to
determine what this quantity corresponds to in the dual CFT. By
construction for $d=4$, this is precisely the central charge $a$.
Motivated by Cardy's general conjecture for even $d$, it is natural to
compare $\ads$ to the coefficient $A$ in eq.~\reef{trace}. In fact,
using the approach of \cite{adam}, one readily confirms that there is a
precise match:
\be \ads =  A \quad{\rm for\ even\ }d\,. \labell{evend} \ee
Hence again, we find support for Cardy's conjecture with this broad
class of holographic CFT's. However, we must seek a broader definition
of $\ads$ to also incorporate odd $d$.

Examining the results for black hole entropy in quasi-topological
gravity \cite{old1}, we observe the following: If the dual CFT is
placed on a ($d\!-\!1$)-dimensional hyperbolic plane (\ie the
$d$-dimensional space is $H^{d-1}\times R$), the energy density
$\rho_\mt{E}$ of the ground state is negative. If a temperature is
introduced and the system is heated up to the point where $\pe=0$, the
entropy density becomes
\be s={(4\pi)^{d/2}}\Gamma\left(d/2\right)\,\ads\ T^{d-1}={2\pi\over
\pi^{d/2}}\Gamma\left(d/2\right)\,{\ads\over L^{d-1}}\,, \labell{bhs}
\ee
where $L$ denotes the radius curvature of $H^{d-1}$. Now this leads to
an interesting question: The bulk geometry is precisely AdS$_{d+1}$ in
`unusual' coordinates, \ie
\be ds^2={dr^2\over \left({r^2\over\tilde L^2}-1\right)}
-\left({r^2\over\tilde L^2}-1\right)\,dt^2 + r^2\,d\Sigma^{d-1}_2
\labell{metric1} \ee
%
\begin{figure}[t]
\begin{tabular}{r}
\includegraphics[width=.4\textwidth, bb=140 320 528 601, clip=true]{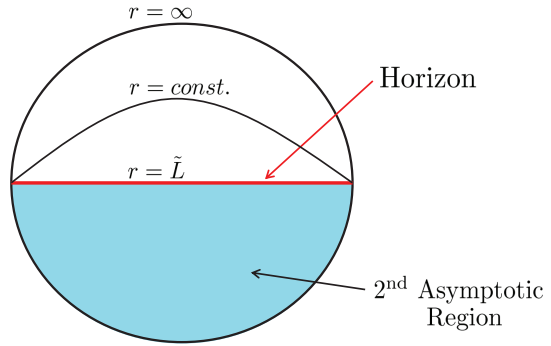}\\
\end{tabular}
\caption{A slice of constant $t$ through the AdS$_{d+1}$ metric in
eq.~\reef{metric1}. This slice bears some similarity to the
Einstein-Rosen bridge in a Schwarzschild black hole
\cite{HE}. Note that only half of the AdS boundary is
reached in the limit $r\rightarrow\infty$. The other half is reached
from the second asymptotic region `behind the horizon'.} \label{picture}
\end{figure}
where $d\Sigma^{d-1}$ denotes the line element on $H^{d-1}$ with unit
curvature, and so why should there be any entropy at all? The answer is
that the foliation of AdS$_{d+1}$ in eq.~\reef{metric1} divides the
boundary into two halves -- see figure \ref{picture}. Hence the entropy
is interpreted as the entanglement entropy between these two halves --
related ideas were discussed in \cite{roberto}. Note that if we
integrate over the hyperbolic horizon out to some maximum radius which
is then given the conventional interpretation of a short-distance
cut-off, \ie $\delta=L^2/\rho_{max} $, the leading contribution in
eq.~\reef{bhs} takes the form $S\propto \ads\,(L/\delta)^{d-2}$.
Hence despite doing a ($d\!-\!1$)-dimensional integral, the leading
divergence produced by the hyperbolic geometry has precisely the power
expected for the `area law' contribution to the entanglement entropy in
a $d$-dimensional CFT \cite{taka}. However, this divergent contribution
is not universal. Rather a universal contribution is extracted from the
subleading terms. The form of the universal contribution to the
entanglement entropy depends on whether $d$ is odd or even:
\be S_{univ}=\left\lbrace
\begin{matrix}
(-)^{\frac{d}{2}-1}\, {4}
\, \ads\, \log(L/\delta)&\quad&{\rm for\ even\ }d\,,\\
(-)^{\frac{d-1}{2}}\,\, {2\pi} \, \ads\ \ \ \ \ \ \ \ \ \ \ &\quad&{\rm
for\ odd\ }d\,.
\end{matrix}\right.
\labell{unis} \ee
With a Weyl transformation, the boundary metric can be brought to that
on $S^{d-1}\times R$ and eq.~\reef{unis} can then be interpreted as the
universal contribution to the entanglement entropy between the two
halves of the $S^{d-1}$ for ground state of the CFT. Hence this
framework allows us to interpret the coefficient $\ads$ in terms of
entanglement entropy in the dual CFT for odd or even $d$. Further our
previous analysis shows that this coefficient obeys a c-theorem in the
holographic RG flows.

\noindent {\bf Discussion:} Having identified this interesting
behaviour in the RG flows of a broad class of holographic CFT's, it is
natural to conjecture that the same c-theorem should apply more broadly
for RG flows outside of a holographic framework. Hence we propose the
general conjecture presented above in the introduction. Now one would
like to determine what evidence one might find for this general
conjecture, again, outside of a holographic framework.

For even dimensions, one can calculate the entanglement entropy for a
CFT divided by a smooth boundary making use of the trace anomaly
\cite{taka,solo}. In general, the result will depend on all of the
coefficients appearing in eq.~\reef{trace}. However, our conjecture
refers to a very specific background geometry and a specific boundary
dividing this space. Applying the approach of \cite{taka,solo} for this
geometry in arbitrary even $d$, one finds that the coefficient of the
universal entanglement entropy \reef{unis} is precisely $\ads=A$
\cite{future}. While this matches eq.~\reef{evend} for our holographic
model, our result here is a general statement about $S_{univ}$ on
$S^{d-1}\times R$ with any CFT in even $d$. Hence our conjectured
c-theorem coincides precisely with Cardy's proposal for arbitrary even
$d$. Therefore any evidence for Cardy's conjecture also supports the
present conjecture. However, the present results also frame Cardy's
conjecture in the context of entanglement entropy. This new perspective
may prove useful in better understanding the precise conditions under
which $A$ satisfies eq.~\reef{beta} in higher (even) $d$.

We emphasize that specifying the geometry in which one calculates the
entanglement entropy is an important feature of our conjecture. As
noted above, our prescription for the geometry was crucial to have
$S_{univ}\propto A$ in even $d$. We can infer that the entanglement
entropy decreases along RG flows for several known examples in $d=2$
and 3 \cite{vidal2,subir,fradkin}. However, we cannot say that these
examples provide direct support of our conjecture primarily because
they do not concern the precise geometry specified there. We
also anticipate that with further study one can identify other
geometries for which the universal entanglement entropy is proportional
to $\ads$ \cite{future}. For example, $S_{univ}\propto a$ for a $d=4$
CFT when calculated for a spherical boundary in flat space \cite{solo}.

Of course, entanglement entropy has previously been considered in the
context of RG flows and c-theorems. In particular, \cite{casini}
establishes an entropic c-theorem in $d=2$ based purely on
considerations of Lorentz symmetry and the strong subadditivity. This
construction is distinct from our conjecture in $d=2$, where in fact
the latter simply coincides with Zamalodchikov's c-theorem \cite{zam}.

The higher curvature terms play an important role in our holographic
analysis as they allow us to unambiguously identify $\ads$ with the
A-type anomaly in any even $d$. Similarly we are also able to
distinguish $\ads$ from $\hC$, the coefficient appearing in the free
energy density, \ie $f=\hC T^{d}$. This coefficient was also considered
in efforts to extend the c-theorem to higher dimensions
\cite{fradkin2}. In our holographic model,
it does not appear that this coefficient always varies monotonically in
RG flows \cite{future}.

In closing, we note that our calculation of the holographic
entanglement entropy did not make reference to the standard proposal
conjectured by \cite{taka}. We would also note that this is also the
first calculation of such a quantity that includes the contributions of
higher curvature interactions in the bulk gravity theory. However, in
the Einstein gravity limit, \ie $\mu=0=\lambda$, our result \reef{unis}
coincides with that calculated using the standard proposal. The present
calculation may point the way to a more systematic derivation of
holographic entanglement entropy.

\noindent {\bf Acknowledgments:} We thank I.~Affleck, E.~Fradkin,
J.~Gomis, S.~Hartnoll, D.~Kutasov, A.~LeClair, M.~Metlitski, C.~Nayak,
S.~Sachdev, G.~Vidal and X.-G.~Wen for discussions. Research at
Perimeter Institute is supported by the Government of Canada through
Industry Canada and by the Province of Ontario through the Ministry of
Research \& Innovation. RCM also acknowledges support from an NSERC
Discovery grant and funding from the Canadian Institute for Advanced
Research.


\begin{thebibliography}{99}
\bibitem 
 {convention} Our convention is that $T_{ab}\equiv-2/\sqrt{-g}\,\delta I/\delta g^{ab}$.
The Euler density $E_d$ is normalized so that on a $d$-dimensional
sphere: $\oint_{S^d}d^d\!x\sqrt{g}\, E_d =2$. The Weyl invariants $I_i$
are constructed from contractions of $d/2$ curvatures or $d/2-k$
curvatures and $2k$ covariant derivatives. Further, note that we have
ignored a scheme-dependent total derivative which may appear in
eq.~\reef{trace}.
\bibitem 
{avatar} For 
reasons explained in \cite{old1}, one should set $\mu=0$ in $d=5$ which
avoids the singularity in last term of eq.~\reef{adfun}.
\end{thebibliography}
\end{document}